\documentclass[10pt,fleqn]{article}
\usepackage{amsfonts,amssymb}
\usepackage{amsmath} %\pagestyle{empty}
\usepackage{graphicx}
\allowdisplaybreaks[1]

\let\kappa=\varkappa

\usepackage[text={17cm,24cm},centering]{geometry}

\newcommand{\dd}{\;\mathrm{d}}
\newcommand{\deriv}[2]{\mbox{$\displaystyle \frac{{\rm d}#1}{{\rm d}#2}$}}
\newcommand{\pderiv}[2]{\,\mbox{$\displaystyle\dfrac{{\partial}#1}{{\partial}#2}$}\,}

\newcommand{\gn}{\gamma_{\mathrm{N}}}

\newcommand{\bm}[1]{\boldsymbol{#1}}
\newcommand{\abc}[1]{\mbox{#1)}\quad}
\newcommand{\pot}[1]{\cdot 10^{#1}}
\newcommand{\kk}[1]{\,\mathrm{#1}}

\newcommand{\ci}[1]{\stackrel{\circ}{#1}{\!\!}}

\newcommand{\vS}{\varSigma}

\newenvironment{Literatur}{\begin{list}{}{\parsep 2pt plus1pt minus1pt
\itemsep \parsep \leftmargin4em\itemindent-4em}}{\end{list}}

\let\Phi=\varPhi
\begin{document}
\title{%
New model of angular momentum transfer from the rotating central
body of a two-body system into the orbital motion of this system
(with application to the earth-moon system)}

\author{        E.Schmutzer, Jena, Germany  \\
         Friedrich Schiller University Jena}
\date{}

\maketitle
\begin{abstract}
In a previous paper we treated within the framework of our
Projective Unified Field Theory (Schmutzer 2004, Schmutzer 2005a)
the 2-body system (e.g. earth-moon system) with a rotating central
body in a rather abstract manner. Here a concrete model of the
transfer of angular momentum from the rotating central body to the
orbital motion of the whole 2-body system is presented, where
particularly the transfer is caused by the inhomogeneous
gravitational force of the moon acting on the oceanic waters of the
earth, being modeled by a spherical shell around the solid earth.
The theory is numerically
tested.\\[1ex]
\end{abstract}

\section{Introduction}\label{sec:1}

Nowadays the concept that the tidal braking effect of the earth
rotation is caused by the gravitational force of the moon acting on
the viscous oceanic waters around the earth is accepted in general.
With respect to the details of this braking mechanism several models
were proposed and theoretically investigated in detail, including
numerical computer calculation. We refrain from reviewing the
historically first ideas and the afterwards following  publications
on this subject, but we rather cite some previous papers, where the
interested reader can find further information: Brosche 1975, 1979,
1989 as well as Brosche and Schuh 1998.

\begin{figure}
\parbox[t]{0.45\textwidth}{\includegraphics[width=0.45\textwidth]{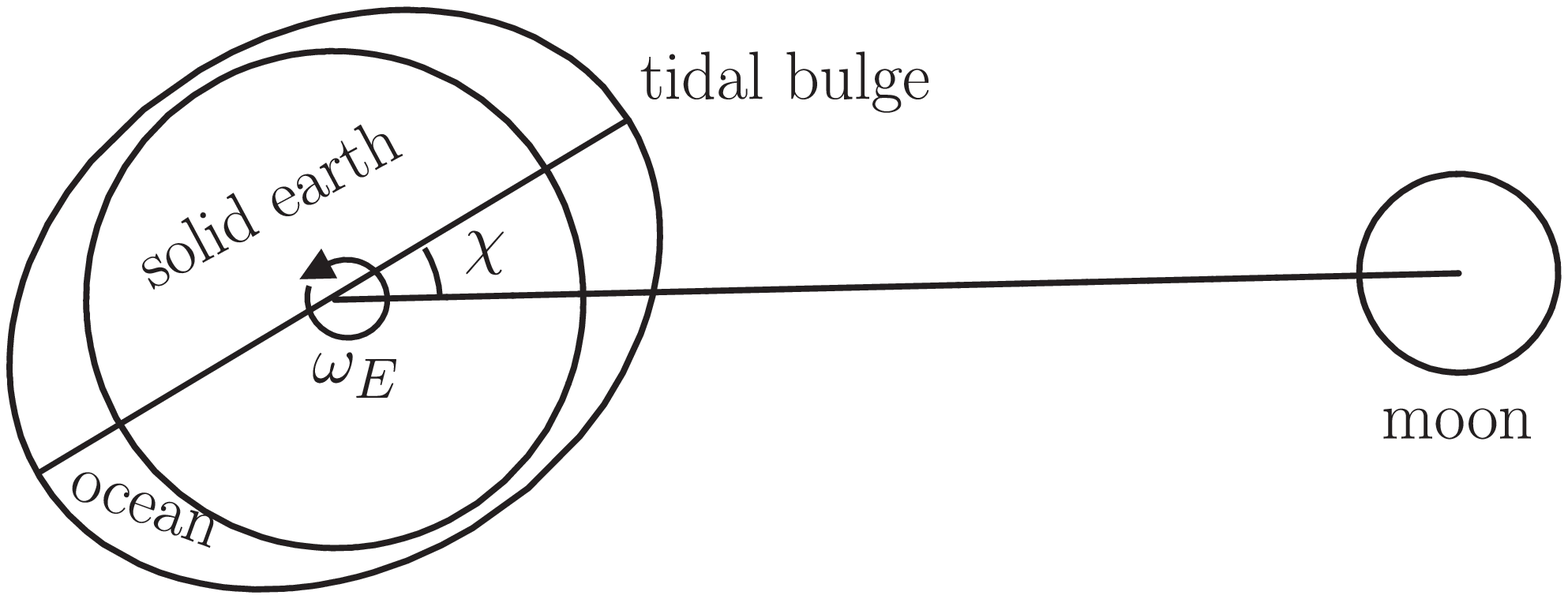}
  \caption{Deformed fluid shell (deformation caused by the
  gravitational field of the moon) including the rotation of the
  earth, which causes the phase lag by the dragging effect within
  the viscous fluid.}\label{Pic1}}\hfill
  \parbox[t]{0.45\textwidth}{
  \includegraphics[width=0.4\textwidth]{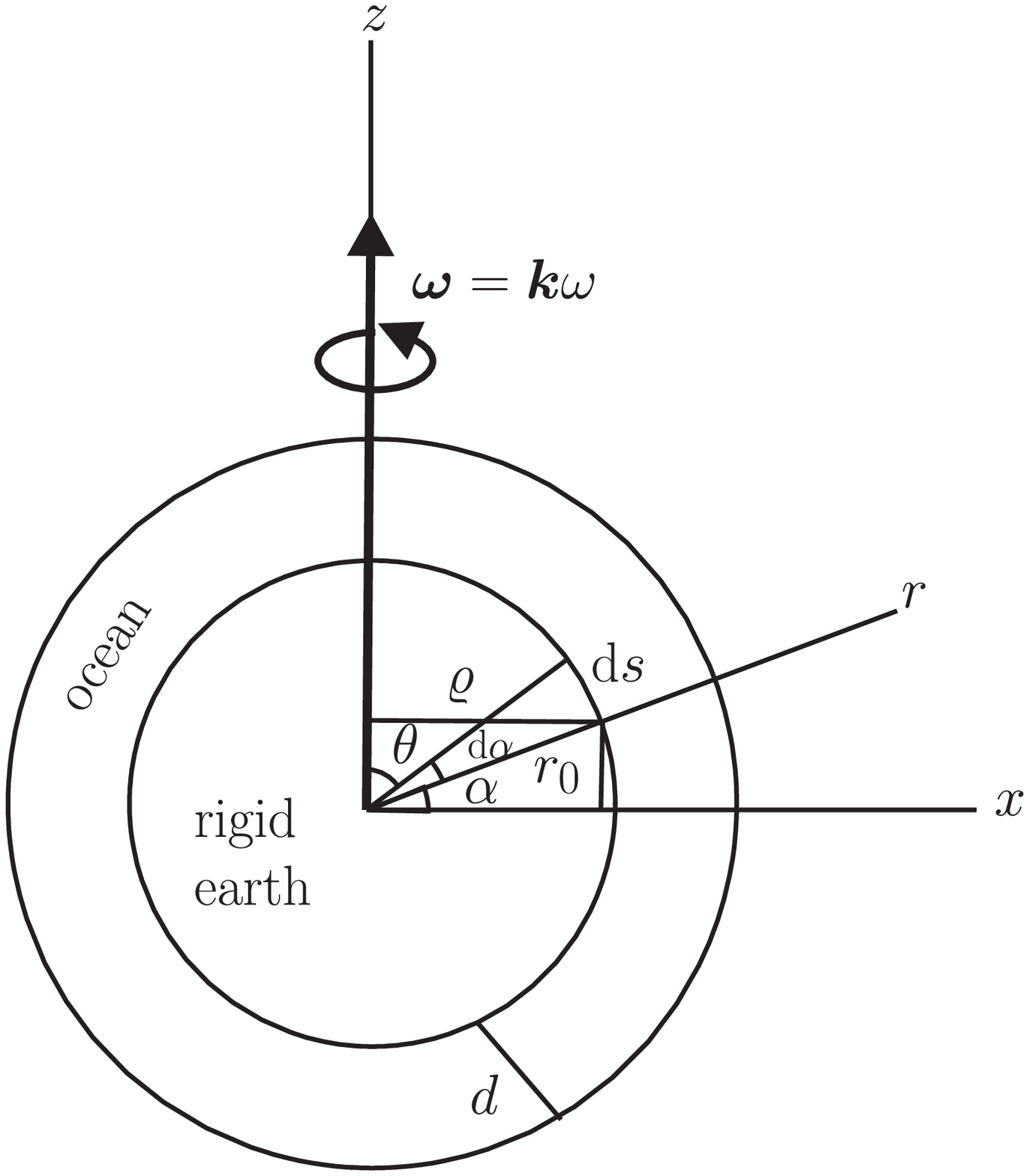}
  \caption{Rigid earth, ocean as viscous fluid shell and line
element $\dd s$ at the interface between both  parts of the earth
($r_0$ radius of the rigid earth).}\label{Pic2}}
\end{figure}

Our considerations start with a strongly simplified model of the
earth consisting of a rigid sphere surrounded by a homogeneous
viscous fluid shell (oceanic waters) of uniform thickness, being
under the influence of  the rotation and the gravitational field of
the rigid earth as well as of the gravitational field of the
orbiting moon. It may be a good help to the reader to have a look on
the figures of forming of the tidal bulge mechanism without and with
phase lag in both the monographs: Lambeck 1980, Sabadini and
Vermeersen 2004. For this publication the physical situation of the
earth-moon system with phase lag is sketched in Fig. \ref{Pic1}.

\begin{figure}
  \includegraphics[width=0.85\textwidth]{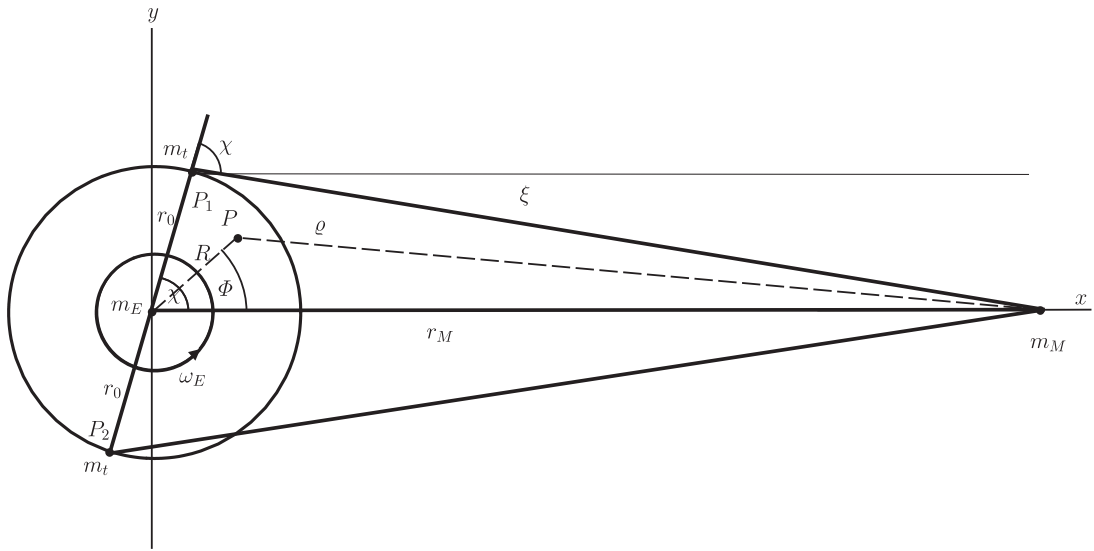}
  \caption{Equilibrium of the tidal bulge (quasi-mechanical body
with mass $m_t$ in the point $P_1$) as a result of the acting
gravitational force of the moon and the dragging force (negative
friction force) within the fluid, caused by rotation of the
earth.}\label{Pic3}
\end{figure}

For the concrete calculations in the following sections we use Fig.
\ref{Pic2} (axis of rotation $\rightarrow$ $z$-axis) and Fig.
\ref{Pic3} (axis of rotation $\rightarrow$ perpendicular to the
$x$-$y$-plane).

\section{Friction force and friction torque}\label{sec:2}

Now using Fig. 2, we apply Newton's friction force law to the
infinitesimal azimuthal component of the friction force $\dd
F_{fric|\Phi}$ within a viscous fluid ($\eta$ viscosity), referring
to an infinitesimal areal element $\dd A_{\Phi}$ positioned at the
point with the cylindrical coordinates ($\varrho=r_0\cos \alpha$,
$\Phi$, $z=r_0\sin\alpha$) at the bottom of the ocean:
\begin{align}\label{1}
\begin{aligned}
    &\abc{a}\dd F_{fric|\Phi}=\eta\deriv{v}{r}\,\dd
    A_{\Phi}=r_0^2\deriv{v}{r}\,
    \cos\alpha\dd \Phi\dd\alpha\quad\text{with}\\
    &\abc{b}\dd A_{\Phi}=\varrho\dd \Phi\dd s\,,\quad
    \abc{c}\dd s=r_0\dd\alpha\,,\quad
    \abc{d}
    v=\omega\varrho=r_0\omega\cos\alpha
\end{aligned}
\end{align}
($r_0$ radius of the earth, $v$ radial-dependent azimuthal velocity,
$\omega$ rotational angular velocity).

In vectorial writing the corresponding infinitesimal friction force
reads:
\begin{align}\label{2}
    \dd \bm{F}_{fric|\Phi}=\bm{e}_{\Phi}\dd F_{fric|\Phi}\,.
\end{align}
Because of the rotation-symmetry assumed we find by integration over
the azimuthal angle $\Phi$ from $0$ to $2\pi$ the intermediate
results
\begin{align}\label{3}
      \abc{a}
      \dd A=\int\limits_{0}^{2\pi}\dd
      A_{\Phi}=2\pi r_0^2\cos\alpha\dd \alpha\,,\quad
      \abc{b}\dd F_{fric}=\eta\deriv{v}{r}\,\dd A\,.
\end{align}
For simplification of the problem we suppose a linear radial
velocity profile ($d$ depth of the ocean):
\begin{align}\label{4}
    \abc{a}
    v(r,\alpha)=v(r_0,\alpha)\Bigl[1-\frac{1-s_{oc}}{d}\,(r-r_0)\Bigr]\,,
    \quad\text{i.e.}\quad
    \abc{b}
    \deriv{v(r,\alpha)}{r}=-\frac{1-s_{oc}}{d}\,v(r_0,\alpha)\,,
\end{align}
where
\begin{align}\label{5}
    v_{d}(\alpha)=s_{oc}v(r_0,\alpha)
\end{align}
is the velocity at the surface of the ocean. We name the constant
$s_{oc}<1$ ``oceanic flow slope parameter''.

Eliminating the velocity gradient in (\ref{3}b), we find
\begin{align}\label{6}
    \dd F_{fric}=-\frac{\eta(1-s_{oc})v(r_0,\alpha)}{d}\,\dd A
\end{align}
and further with the help of (\ref{3}a) and (\ref{1}d)
\begin{align}\label{7}
    \dd F_{fric}=-\frac{2\pi
    r_0^3\omega\eta(1-s_{oc})\cos^2\alpha}{d}\,\dd \alpha\,.
\end{align}
Integration over $\alpha$ from $-\dfrac{\pi}{2}$ to $\dfrac{\pi}{2}$
yields
\begin{align}\label{8}
    F_{fric}=-\frac{\pi^2r^3_0\omega\eta(1-s_{oc})}{d}\,.
\end{align}
Adapting this uniform oceanic shell modeling to the earth, where
about 70.8 \% of the surface are oceanic waters, it is convenient to
introduce a rough ``surface correction factor'' $f_{oc}$. Then
formula (\ref{8}) reads
\begin{align}\label{9}
    F_{fric}=-\frac{\pi^2r^3_0\omega\eta(1-s_{oc})f_{oc}}{d}\,.
\end{align}
As an example one is tempted to choose $f_{oc}\approx 0.6$.

Now we define in the usual way the infinitesimal vectorial friction
torque
\begin{align}\label{10}
    \dd
    \bm{M}_{fric|\Phi}=\bm{e}_{r}r_0\times\dd\bm{F}_{fric|\Phi}=-\bm{e}_{\theta}r_0\dd
    F_{fric|\Phi}
\end{align}
corresponding to the infinitesimal friction force (\ref{2}).

In analogy to the above calculations we receive by integration
\begin{align}\label{11}
    \dd \bm{M}_{fric}=\int\limits_{\Phi=0}^{2\pi}\dd
    \bm{M}_{fric|\Phi}\dd \Phi=
    -\bm{k}\frac{2\pi
    r_0^4\omega\eta(1-s_{oc})\cos^3\alpha\dd\alpha}{d}
\end{align}
and by further integration
\begin{align}\label{12}
    \bm{M}_{fric}=-\bm{k}\frac{2\pi
    r_0^4\omega\eta(1-s_{oc})f_{oc}}{d}\!
    \int\limits_{\alpha=-\frac{\pi}{2}}^{\frac{\pi}{2}}\cos^3\alpha\dd\alpha
\end{align}
or  explicitly
\begin{align}\label{13}
    \abc{a}
    \bm{M}_{fric}=\bm{k}M_{fric}\quad\text{with}\quad\abc{b}
    M_{fric}=-\frac{8\pi r_0^4\omega\eta(1-s_{oc})f_{oc}}{3d}\,.
\end{align}
Let us now remember some results from our previous paper (Schmutzer
2005a): We used the following definition of the rotational angular
momentum of the earth rotating about the $z$-axis:
\begin{align}\label{14}
    \bm{L}=\bm{k}L_{rot}=\bm{k}\theta_E\omega_E
\end{align}
($\theta_E$ moment of inertia, $\omega_E$ rotational angular
velocity, $\bm{k}$ unit vector in $z$-direction). Hence by temporal
differentiation follows (dot means total time derivative)
\begin{align}\label{15}
 \abc{a}   \deriv{\bm{L}}{t}=\bm{k}\bigl(\dot{\theta}_{E}\omega_E+\theta_E\dot{\omega}_E\bigr)\quad
    \text{with}\quad
  \abc{b}\dot{\theta}_E={}\ci{\theta}_E+\theta_E\vS\,.
\end{align}
From this last equation we remember that the total time derivative
of the moment of inertia consists of two parts: the partial
derivative (denoted by circle) and the temporal cosmological
influence (determined by the ``logarithmic scalaric world function''
$\vS$).

According to the balance equation of the angular momentum
\begin{align}\label{16}
    \deriv{\bm{L}}{t}=\bm{M}_{fric}
\end{align}
the temporal change of $\bm{L}$ is caused by the friction torque
(\ref{13}a). Inserting the expression (\ref{15}a) and (\ref{13}a)
into (\ref{16}) yields the following temporal change of the angular
velocity (for $\omega\to\omega_E$):
\begin{subequations}\label{17}
\begin{align}\label{17a}
&\dot{\omega}_E=-\frac{8\pi
r_0^4\omega_E\eta(1-s_{oc})f_{oc}}{3\theta_E
d}-\frac{(\ci{\theta}_E+\theta_E\vS)\omega_E}{\theta_E}\quad
\text{or}\\
\label{17b} &\frac{\eta(1-s_{oc})f_{oc}}{d}=
-\frac{3\bigl[\theta_E\dot{\omega}_E+(\ci{\theta}_{E}+\theta_E\vS)\omega_E\bigr]}{8\pi
r_0^4\omega_E}\,.
\end{align}
\end{subequations}

\section{Deformation of the earth by the inhomogeneous gravitational field of the
moon}

Basis of our following considerations is Fig. \ref{pic3}. For
treating the above physical problem further we first calculate the
rotation-symmetric gravitational field of the moon (taken as
point-like) about the $x$-axis ($R=\sqrt{x^2+y^2}$) under the
approximate suppositions:
\begin{align}\label{18}
    \abc{a} \frac{r}{r_M}\ll
    1\,,\quad\abc{b}\frac{m_Mr_0^2}{m_Er_M^2}\ll1\,,
    \quad\abc{c}\frac{m_Mr_0^3}{m_Er_M^3}\ll 1
\end{align}
($m_E$ mass of the earth, $m_{M}$ mass of the moon, $r_M$ distance
between the centers of earth and moon).

First we remember that the gravitational force of the moon acting on
the center of the earth is compensated by the centrifugal force in
this same point, caused by the orbital motion of the earth about the
common center of mass of the 2-body problem.

Now we begin our considerations on the gravitational potential
$\phi_M$ of the moon at an arbitrary point $P$  inside the earth
($r\leqq r_0$). Using the notation of  Fig. \ref{pic3}, then the
well-known Newtonian potential reads
\begin{align}\label{19}
    \phi_M(\varrho)=-\frac{\gn m_M}{\varrho}\,.
\end{align}
Hence for the center of the earth results
\begin{align}\label{20}
    \phi_M(r_M)=-\frac{\gn m_M}{r_M}\,.
\end{align}
The difference of both the potentials is equal to
\begin{align}\label{21}
    \varphi_M(\varrho)=\phi_M(\varrho)-\phi_M(r_M)=-\gn
    m_M\Bigl(\frac{1}{\varrho}-\frac{1}{r_M}\Bigr)\,.
\end{align}
According to the cosine law we learn from Fig. \ref{pic3}:
\begin{align}\label{22}
    \varrho=
    r_M\Bigl[1+\Bigl(\frac{R}{r_M}\Bigr)^2-\frac{2R}{r_M}\,\cos\Phi\Bigr]^{1/2}\,.
\end{align}
With the help of (\ref{18}a) we approximately find
\begin{align}\label{23}
    \varrho=
    r_M\Bigl[1-\frac{R}{r_M}\,\cos\Phi+\frac{1}{2}\,\Bigl(\frac{R}{r_M}\Bigr)^2\Bigl(1-\cos^2\Phi\Bigr)\Bigr]
\end{align}
and further
\begin{align}\label{24}
    \varphi_{M}(\varrho)=-\frac{\gn m_M}{r_M}\,\Bigl[
    \frac{R}{r_M}\,\cos
    \Phi-\frac{1}{2}\Bigl(\frac{R}{r_M}\Bigr)^2(1-3\cos^2\Phi)\Bigr]\,.
\end{align}
Rearranging this equation gives
\begin{align}\label{25}
    \abc{a}\varphi_M(x,y)=\frac{\gn
    m_M}{2r_M^3}\,(y^2-2x^2-2xr_M)\quad\text{with}\quad
    \abc{b}R^2=x^2+y^2\,.
\end{align}
With respect to the earth, for simplicity we refer to a homogeneous
mass sphere. From usual textbooks we take the expression for the
gravitational potential in the interior of such a homogeneous
spherical body (in this notation e.g. Schmutzer 2005b):
\begin{subequations}\label{26}
    \begin{align}\label{26a}
&\phi_{Ei}(r)=\frac{\gn m_E}{2r_0^3}\,(r^2-3r_0^3) &&
\text{(interior potential),}\\
& \phi_{Ee}(r)=-\frac{\gn m_E}{r} && \text{(exterior potential).}
\end{align}
\end{subequations}
From (\ref{25}a) and (\ref{26}a) we receive the superposition
potential of the earth and the moon in the interior of the earth:
\begin{align}\label{27}
    \phi_i(x,y)=\phi_{Ei}(x,y)+\varphi_M(x,y)=
    \frac{\gn
    m_E}{2r_0}\,\Bigl[\Bigl(1-\frac{2m_Mr_0^3}{m_Er_M^3}\Bigr)\frac{(x-x_0)^2}{r_0^2}+
    \Bigl(1+\frac{m_M r_0^3}{m_E
    r_M^3}\Bigr)\frac{y^2}{r_0^2}\Bigr]-\phi_0\,,
\end{align}
where the abbreviations
\begin{align}\label{28}
    \abc{a}\phi_0=\frac{\gn
    m_M^2r_0^3}{2m_Er_M^4\Bigl(1-\dfrac{2m_Mr_0^3}{m_Er_M^3}\Bigr)}+
    \frac{3\gn m_E}{2r_0}\,,\quad
    \abc{b}x_0=\frac{m_Mr_0^3}{m_Er^2_M\Bigl(1-\dfrac{2m_Mr_0^3}{m_Er_M^3}\Bigr)}
\end{align}
were used.

Let us here mention that the treatment of this subject in the
corresponding literature usually starts from the integral
representation of the potential. Our superposition method applied
above lead us to the resulting ellipsoid formula by setting
$\phi_i(x,y)=\mathrm{const}$ in (\ref{27}). This approach shows that
under the influence of the moon, beyond the deformation of the
sphere to an ellipsoid, our calculation exhibits a translation of
the resulting ellipsoid by the amount $x_0$ in the $x$-direction.

Using the semi-axes $\hat{a}$ and $\hat{b}$, taken from the above
abbreviations,
\begin{align}\label{29}
    \abc{a}\Hat{a}^2=\frac{r_0^2}{1-\dfrac{2m_Mr_0^3}{m_Er_M^3}}\,,
    \quad
    \abc{b}
    \Hat{b}^2=\frac{r_0^2}{1+\dfrac{m_M r_0^3}{m_Er_M^3}}\,,
\end{align}
the equation (\ref{27}) reads
\begin{align}\label{30}
    \phi_i(x,y)=\frac{\gn
    m_E}{2r_0}\Bigl[\frac{(x-x_0)^2}{\hat{a}^2}+\frac{y^2}{{\hat{b}}^2}-1\Bigr]+\frac{\gn
    m_E}{2r_0}-\phi_0\,.
\end{align}
Fitting this potential to the surface equation of the physical
ellipsoid
\begin{align}\label{31}
    \frac{(x-x_0)^2}{\hat{a}^2}+\frac{y^2}{{\hat{b}}^2}=1
\end{align}
of the slightly deformed earth by the inhomogeneous gravitational
field of the moon leads to the expression
\begin{align}\label{32}
    \phi_i(x,y)_{\mathit{ellipsoid}}=\frac{\gn
    m_E}{2r_0}-\phi_0=-\frac{\gn m_E}{r_0}-
    \frac{\gn m_M^2r_0^3}{2m_Er^4_M\Bigl(1-\dfrac{2m_M r_0^3}{m_E
    r_M^3}\Bigr)}\,.
\end{align}
Further simplification of the above formula (\ref{31}) is reached by
applying the approximate suppositions derived from (\ref{18}):
\begin{align}\label{33}
    \abc{a}x_0=\frac{m_Mr_0^3}{m_Er_M^2}\,,\quad
    \abc{b}\phi_0=\frac{3\gn m_E}{2r_0}\,,
    \quad
    \abc{c}\hat{a}=r_0\Bigl(1+\frac{m_Mr_0^3}{m_Er_M^3}\Bigr)\,,
    \quad
    \abc{d}
    \hat{b}=r_0\Bigl(1-\frac{m_Mr_0^3}{2m_Mr_M^3}\Bigr)\,.
\end{align}
Conservation of the mass of the earth
\begin{align}\label{34}
    m_E=\frac{4\pi}{3}\,\mu_0r_0^3
\end{align}
($\mu_0$ constant mass density) during the deformation leads to the
relation
\begin{align}\label{35}
    r_0^3=\hat{a}{\hat{b}}^2\,.
\end{align}
One should remember that for simplicity our above calculations were
based on a homogeneous mass model of the earth. For our further
calculations we are only interested in the physical situation at the
surface of the sphere, not explicitly referring to its interior.
Therefore our further treatment is consistent.

\section{Rotation of the earth, dragging force on the tidal bulges and
calculation of the tidal phase lag angle (tidal dragging
angle)}\label{sec:4}

\subsection{Forces on the moon-nearest bulge}\label{sec:4.1}

Nowadays there is no doubt among the experts that the phenomenon of
the oceanic tides is primarily caused by the moon. The observed
tidal water bulges in the zenith (direction to the moon) and
symmetric to it in the nadir (opposite direction) are modeled by the
two water bulges resulting from the symmetric deformation of the
sphere to the ellipsoidal form calculated above. Since the
rotational angular velocity of the earth is much bigger than the
revolution angular velocity of the moon, the earth performs a
relatively quick rotation with respect to the tidal bulges, i.e.
with respect to the moon (motion under the quasi-fixed bulges).
Hence follows that an observer in the frame of reference of the
surface of the earth realizes the phenomenon of tides.

Here we investigate the dragging effect of the viscous waters,
caused in the waters by the rotation of the rigid earth. This effect
leads to a shift of the bulges by a certain phase lag angle $\chi$.
The equilibrium position of the bulges is the result of two on the
bulges acting torques, caused by the gravitational force of the moon
and the dragging force in the viscous oceanic waters, as above
already mentioned.

Our task is now to determine this equilibrium position of the
moon-nearest bulge which abstractively will be considered as a
mechanical quasi-body, where the forces act on the center of mass of
this quasi-body ($m_t$ tidally caused mass of the quasi-body). The
following investigations are based on Fig. \ref{pic3}, where for
simplicity the ellipsoid is approximated by a sphere and the point
of consideration is shifted from the interior to the surface of the
earth ($P\to P_1$).

Let us further mention that the gravitational force of the sun,
which is distinctly smaller than the gravitational force of the
moon, will be neglected for simplification of the physical problem,
i.e. we concentrate our investigation on the primarily essential
points of the equilibrium  mechanism.

In order to calculate the resulting torque acting on the center of
mass of the moon-nearest bulge, we first list the various forces
being present:
\begin{align}\label{36}
    \bm{F}_E=m_t \bm{G}_E=m_t\bm{e}_RG_{E|R}
\end{align}
(radial gravitational force of the earth),
\begin{align}
\label{37}
   \bm{F}_M=m_t\bm{G}_M
    \end{align}
(radial gravitational force of the
    moon),
\begin{align}
     \label{38}
    \bm{F}_c=m_t\bm{G}_c
=m_t\bm{e}_{R}r_0\omega_E^2
\end{align}
(radial centrifugal force by rotation of the earth),
\begin{align}
    \label{39}
\bm{F}_{fric}=\bm{e}_{\Phi}F_{fric}
\end{align}
(azimuthal friction force determined by (\ref{9})),
\begin{align}\label{40}
    \bm{F}_p=\bm{e}_RF_{p|R}
\end{align}
(radial pressure force on the bulge as back-reaction of the earth).

One should realize that the quantities $\bm{G}_{E}$,  $\bm{G}_{M}$
and $\bm{G}_{c}$ are acceleration quantities.

Using the radial and azimuthal vectorial decomposition of
(\ref{37}), we obtain
\begin{align}\label{41}
    \abc{a}\bm{F}_M=\bm{e}_RF_{M|R}+\bm{e}_\Phi
    F_{M|\Phi}\quad\text{with}\quad\abc{b}
    F_{M|R}=m_tG_{M|R}\,,\quad\abc{c}
    F_{M|\Phi}=m_tG_{M|\Phi}\,.
\end{align}
Hence follows the radial component of the total force as the sum of
all radial components:
\begin{align}\label{42}
    F_R=m_t\bigl(G_{E|R}+G_{M|R}+r_0\omega_E^2\bigr)+F_{p|R}\,.
\end{align}

\subsection{Stationary equilibrium}\label{sec:4.2}

Demanding stationarity we arrive at the equation
\begin{align}\label{43}
    F_{R}=0\,.
\end{align}
With respect to the azimuthal component of the force we meet the
following physical situation: Due to the viscosity of the fluid the
quasi-body (modeled bulge) is dragged into the direction of the
rotational orbital velocity by the dragging force
\begin{align}\label{44}
    F_{drag}=-F_{fric}\qquad \text{($F_{drag}>0$, $F_{fric}<0$).}
\end{align}
In the case of stationarity the dragging force is compensated by the
(oppositely directed) gravitational force (pulling force) of the
moon.

These previous considerations lead us to the equilibrium equation
\begin{align}\label{45}
    F_{\Phi}=F_{M|\Phi}+F_{drag}=m_tG_{M|\Phi}+F_{drag}=0
\end{align}
being by means of (\ref{44}) equivalent to the condition
\begin{align}\label{46}
    F_{fric}=m_tG_{M|\Phi}\,.
\end{align}
According to the definition of the torque by the azimuthal forces
\begin{align}\label{47}
    M_{\Phi}=F_{\Phi}r_0\,,
\end{align}
for the equilibrium results
\begin{align}\label{48}
    M_{\Phi}=0\,,
\end{align}
being equivalent to (\ref{45}).

\subsection{Calculation of the azimuthal gravitational force component
and of the gravitational torque, both caused by the
moon}\label{sec4:3}

After these decomposition procedures we are now able to calculate on
the basis of  Fig. \ref{pic3} the azimuthal component of the
gravitational force of the moon acting on the tidal bulge being at
the oceanic surface of the earth ($R\to r_0$, $\Phi\to \chi$). In
this specialization formula (\ref{23}) reads
\begin{align}\label{49}
    \xi
    =r_M\Bigl[1-\frac{r_0}{r_M}\,\cos\chi+\frac{1}{2}\,\Bigl(\frac{r_0}{r_M}\Bigr)^2(1-\cos^2\chi)\Bigr]\,.
\end{align}
By differentiation follows
\begin{align}\label{50}
    \pderiv{\xi}{\chi}=r_0\sin\chi
    \Bigl(1+\frac{r_0}{r_M}\,\cos\chi\Bigr)\,.
\end{align}
We further receive with the help of (\ref{49}) from (\ref{19})
\begin{align}\label{51}
    \phi_M(\xi)=-\frac{\gn
    m_M}{r_M}\,\Bigl[1+\frac{r_0}{r_M}\,\cos\chi-\frac{1}{2}\,\Bigl(\frac{r_0}{r_M}\Bigr)^2(1-3\cos^2\chi)\Bigr]
\end{align}
and hence by differentiation
\begin{align}\label{52}
    \pderiv{\phi_M(\xi)}{\xi}=\frac{\gn
    m_M}{r_M^2}\,\Bigl[1+\frac{2r_0}{r_M}\,\cos\chi-\Bigl(\frac{r_0}{r_M}\Bigr)^2(1-4\cos^2\chi)\Bigr]
    \,.
\end{align}
By means of (\ref{52}) and (\ref{50}) follows
\begin{align}\label{53}
    \pderiv{\phi_M}{\chi}=\pderiv{\phi_M(\xi)}{\xi}\pderiv{\xi}{\chi}=
    \frac{\gn
    m_Mr_0\sin\chi}{r_M^2}\,\Bigl[1+\frac{3r_0}{r_M}\,\cos\chi-\Bigl(\frac{r_0}{r_M}\Bigr)^2
    (1-6\cos^2\chi)\Bigr]\,.
\end{align}
This result leads to the azimuthal gravitational force of the moon,
acting on the moon-nearest tidal bulge at the azimuthal angle $\chi$
(first order approximation in $r_0$):
\begin{align}\label{54}
    F_{M\chi}=m_tG_{M\chi}=-m_t\pderiv{\phi_M}{(r_0\chi)}=
    -\frac{\gn
    m_Mm_t\sin\chi}{r_M^2}\,\Bigl(1+\frac{3r_0}{r_M}\,\cos\chi\Bigr)\,.
\end{align}
Hence by the angle transformation $\chi\to \chi+\pi$ we arrive at
the corresponding force on the diagonally situated second tidal
bulge:
\begin{align}\label{55}
    F_{M(\chi+\pi)}=\frac{\gn
    m_Mm_t\sin\chi}{r_M^2}\,\Bigl(1-\frac{3r_0}{r_M}\,\cos\chi\Bigr)\,.
\end{align}
From (\ref{54}) and (\ref{55}) we find for the azimuthal
gravitational force of the moon, acting on both tidal bulges:
\begin{align}\label{56}
    F_{M|az}=F_{M\chi}+F_{M(\chi+\pi)}=-\frac{3\gn
    m_Mm_tr_0}{r_M^3}\,\sin(2\chi)\,.
\end{align}
The corresponding azimuthal gravitational torque, caused by the
moon, reads:
\begin{align}\label{57}
    M_{M|az}=F_{M|az}r_0\,.
\end{align}

\subsection{Tidal phase lag angle (dragging angle)}\label{sce:4.4}

This quantity follows immediately from the stationary equilibrium
condition (\ref{45}). Inserting  (\ref{56}) and (\ref{8}) into this
condition gives the intended result ($\omega\to \omega_E$ angular
velocity of the earth):
\begin{align}\label{58}
    \abc{a}
    \sin(2\chi)=\frac{\pi^2r_0^2r_M^3\omega_E\eta(1-s_{oc})f_{oc}}{3\gn
    m_Mm_t d}\quad\text{or}\quad\abc{b}
    \chi=\frac{1}{2}\,\arcsin\frac{\pi^2r_0^2r_M^3\omega_E\eta(1-s_{oc})f_{oc}}{3\gn
    m_Mm_t d}
\end{align}
($\chi$ tidal phase lag angle or dragging angle).

For practical reasons we approximate (\ref{58}a) for the case
$|\chi|\ll \dfrac{1}{2}$:
\begin{align}\label{59}
    \chi=\frac{\pi^2r_0^2r_M^3\omega_E\eta(1-s_{oc})f_{oc}}{6\gn
    m_Mm_t d}\,.
\end{align}
Further it is useful to introduce the meaningful quantity
\begin{align}\label{60}
    S_{tidal}=\frac{\eta(1-s_{oc})f_{oc}}{d m_t \chi}
\end{align}
which we name ``tidal coefficient''.

By means of (\ref{59}) we receive the following different form of
it:
\begin{align}\label{61}
    S_{tidal}=\frac{6\gn m_M}{\pi^2r_0^2r_M^3\omega_E}\,,
\end{align}
In this context we remember the relation (\ref{17}b). Neglecting the
rebound effect, cosmological effect, etc., from this relation
mentioned we find
\begin{align}\label{62}
    \frac{\eta(1-s_{oc})f_{oc}}{d}=-\frac{3\theta_{E}\dot{\omega}_{E|tidal}}{8\pi
    r_0^4\omega_E}\,.
\end{align}
Eliminating in (\ref{60}) with the help of this formula, we obtain
the following alternative relation for the tidal coefficient:
\begin{align}\label{63}
    S_{tidal}=-\frac{3\theta_E\dot{\omega}_{E|tidal}}{8\pi
    r_0^4\omega_Em_t\chi}\,.
\end{align}

\section{Numerical evaluation}\label{sec:5}

\subsection{Empirical values}\label{sec:5.1}

First from the corresponding literature we list some empirical data
on the earth-moon system etc., being useful for the following
numerical calculations (Gauss system of units):
\begin{align}\label{64}
    \gn=6.673\pot{-8}\kk{g^{-1}cm^3s^{-2}}
\end{align}
(Newtonian gravitational constant),
\begin{align}\label{65}
    r_0=6.378\pot{8}\kk{cm}
\end{align}
(radius of the earth),
\begin{align}\label{66}
    r_M=3.844\pot{10}\kk{cm}
\end{align}
(distance earth/moon),
\begin{align}\label{67}
    m_E=5.976\pot{27}\kk{g}
\end{align}
(mass of the earth),
\begin{align}\label{68}
    m_M=7.348\pot{25}\kk{g}
\end{align}
(mass of the moon).
\begin{align}\label{69}
    \omega_{E}=7.292\pot{-5}\kk{s^{-1}}
\end{align}
(angular velocity of the earth),
\begin{align}\label{70}
    \dot{\omega}_{E|tidal}=-6.15\pot{-22}\kk{s^{-2}}
\end{align}
(tidal braking angular acceleration of the earth),
\begin{align}\label{71}
    \mu_{oc}\approx 1.025\kk{g\,cm^{-3}}\qquad
\text{[IERS Conventions 2003]}
\end{align}
(average mass density of the oceanic waters),
\begin{align}\label{72}
    \theta_E=C_E=8.039\pot{44}\kk{g\,cm^2}
\end{align}
(polar moment of inertia of the earth).

Let us here add two remarks:
\begin{enumerate}
  \item
The values listed above partly serve as data for approximate
information.
    \item
The following numerical calculations are performed for rough testing
our theory with respect to the order of magnitude.
\end{enumerate}

\subsection{Detailed numerical results by using values of the above
list} \label{sec:5.2}

From (\ref{61}) we determine the tidal coefficient:
\begin{align}\label{73}
    S_{tidal}=1.77\pot{-27}\kk{cm^{-2}s^{-1}}\,,
\end{align}
whereas from (\ref{63}) we find
\begin{align}\label{74}
    m_{t}\chi S_{tidal}=4.91\pot{-9}\kk{g\,cm^{-2}s^{-1}}\,.
\end{align}
Using the value (\ref{73}), we arrive at
\begin{align}\label{75}
    m_t\chi=2.77\pot{18}\kk{g}\,.
\end{align}
As empirically observed, the tide is coming in approximately 25
minutes after the culmination of the moon. This fact corresponds to
a phase lag angle of
\begin{align}\label{76}
    \abc{a}
    \chi=6.25^{\circ}=0.11\kk{rad}\,,\quad\text{i.e.}\quad \abc{b}
    \sin\chi=0.112\,.
\end{align}
Inserting the first value into (\ref{75}), we find for the mass of
the tidal bulge the numerical result
\begin{align}\label{77}
    m_t=2.54\pot{19}\kk{g}\,.
\end{align}
Since we here are mainly interested in orders of magnitudes of some
physical quantities, we refrain from refinements of our rough
oceanic shell model. For simplicity we therefore choose for the
thickness of the shell the estimated average value
\begin{align}\label{78}
    d=3.8\pot{5}\kk{cm}=3.8\kk{km}
\end{align}
to be found in literature. Further for our refinement parameters we
take the simple values $s_{oc}=0$, $f_{oc}=1$.

Then from (\ref{60}) results
\begin{align}\label{79}
    \abc{a} S_{tidal}=\frac{\eta}{dm_t\chi}\quad\text{or}\quad
    \abc{b}
    \eta=S_{tidal}dm_t\chi\,.
\end{align}
Using the above numerical values (\ref{73}), (\ref{76}a), (\ref{77})
and (\ref{78}), we arrive at the following approximate value of the
viscosity of the oceanic waters
\begin{align}\label{80}
    \eta=1.87\pot{-3}\kk{g\,cm^{-1}s^{-1}}=1.87\pot{-3}\kk{Poise}\quad
    \text{($1\kk{Poise}=1\kk{g\,cm^{-1}s^{-1}}$)}
\end{align}
which, considering our rough shell model, seems not to be too far
away from empirical viscosity values of water, e.g.
$\eta_{water}\approx 10^{-2}\kk{Poise}$ at $20^{\circ}\kk{C}$.

Finally we estimate the seize of the tidal bulge, up to now roughly
treated as a quasi-mechanical body. We approximate such a body by a
spherical cap whose volume is given by the formula ($R$ radius of
the cap, $h$ height of the cap)
\begin{align}\label{81}
    V_{cap}=\frac{\pi h}{6}\,(3R^2+h^2)\,.
\end{align}
Rearranging this equation leads us to the following formula for the
radius of the cap:
\begin{align}\label{82}
    R=\sqrt{\dfrac{2m_t}{\pi \mu_{oc}h}-\dfrac{1}{3}\,h^2}\,.
\end{align}
Taking from literature the approximate empirical value of the height
of the tide cap $h = 53 \kk{cm}$, by means of (\ref{71}) and
(\ref{77}) we receive for the radius of the cap the numerical value
\begin{align}\label{83}
    R=5.46\pot{8}\kk{cm}=5460\kk{km}\,.
\end{align}

\vspace{1ex}
Thankfully I appreciate interesting discussions with
Prof. P.~Brosche (Observatory Hoher List) and the technical help of
Prof. A.~Gorbatsievich (University of Minsk).

\section*{References}

\begin{Literatur}
  \item
Brosche, P.: 1975, Naturwissenschaften 62, (p.1-9);
             1979, Astron. Nachr. 300, (p.195-196);
             1989, Physik in unserer Zeit (Geophysik), 20.Jahrgang, Nr. 3,
             (p.70-78)
  \item
Brosche, P. , Schuh, H.:  1998, in: Surveys in Geophysics 19,
(p.417-430)
  \item
Lambeck, K.: 1980, The Earth`s Variable Rotation: Geophysical Causes
and Consequences, Cambridge University Press, Cambridge
  \item
Sabadini, R. , Vermeersen, B.: 2004, Global Dynamics of the Earth,
Kluwer Academic Publishers, Dordrecht-Boston-London
  \item
Schmutzer, E.: 2004, Projektive Einheitliche Feldtheorie mit
Anwendungen in Kosmologie und Astrophysik, Verlag Harri Deutsch,
Frankfurt/Main
  \item
Schmutzer, E.: 2005a, Astron. Nachr. (in press)
  \item
Schmutzer, E.: 2005b, Grundlagen der Theoretischen Physik, 3rd
edition, Wiley-VCH, Weinheim ; 1989, Grundlagen der Theoretischen
Physik, 1st edition Deutscher Verlag der Wissenschaften, Berlin

\end{Literatur}

\end{document}